\documentclass[11pt,aps,prb,preprint]{revtex4}
\usepackage{amsmath}
\usepackage{graphicx}
\usepackage{url}

\begin{document}

\title{Increased entropy of signal transduction in the cancer metastasis phenotype}

\author{Andrew E Teschendorff$^*$}
\affiliation{Medical Genomics Group, Paul O'Gorman Building, UCL Cancer Institute, University College London, 72 Huntley Street, London WC1E 6BT, United Kingdom. \\
$^*$Corresponding author: a.teschendorff@ucl.ac.uk}
\author{Simone Severini}
\affiliation{Department of Physics \& Astronomy, University College London, London WC1E 6BT, United Kingdom.}

\begin{abstract}
The statistical study of biological networks has led to important novel biological insights, such as the presence of hubs and hierarchical modularity. There is also a growing interest in studying the statistical properties of networks in the context of cancer genomics. However, relatively little is known as to what network features differ between the cancer and normal cell physiologies, or between different cancer cell phenotypes.\\
Based on the observation that frequent genomic alterations underlie a more aggressive cancer phenotype, we asked if such an effect could be detectable as an increase in the randomness of local gene expression patterns. Using a breast cancer gene expression data set and a model network of protein interactions we derive constrained weighted networks defined by a stochastic information flux matrix reflecting expression correlations between interacting proteins. Based on this stochastic matrix we propose and compute an entropy measure that quantifies the degree of randomness in the local pattern of information flux around single genes. By comparing the local entropies in the non-metastatic versus metastatic breast cancer networks, we here show that breast cancers that metastasize are characterised by a small yet significant increase in the degree of randomness of local expression patterns. We validate this result in three additional breast cancer expression data sets and demonstrate that local entropy better characterises the metastatic phenotype than other non-entropy based measures. We show that increases in entropy can be used to identify genes and signalling pathways implicated in breast cancer metastasis and provide examples of {\it de-novo} discoveries of gene modules with known roles in apoptosis, immune-mediated tumour suppression, cell-cycle and tumour invasion.\\
These results demonstrate that a metastatic cancer phenotype is characterised by an increase in the randomness of the local information flux patterns. Measures of local randomness in integrated protein-interaction mRNA expression networks may therefore be useful for identifying genes and signalling pathways disrupted in one phenotype relative to another. Further exploration of the statistical properties of such integrated cancer expression and protein interaction networks will be a fruitful endeavour.
\end{abstract}
\maketitle

\section*{Background}
The statistical study of biological networks has shed substantial novel insights into how biological networks are organised and how this organisation may relate to cellular function \cite{Barabasi2004}. An important class of biological networks are protein interaction networks (PINs), as determined from high-throughput interaction screens and literature curation \cite{Prasad2009,Stelzl2006,Brown2005}. Edges in these networks mainly describe interactions of proteins that are part of the same physical complex or posttranslational modifications mediating signal transduction flows. Topological studies of these PINs have shown that they have important properties such as scale-freeness of the degree distribution, characterised by most proteins having small degree constrasting with a small yet significant number of highly connected nodes, called hubs \cite{Barabasi2004}. Hubs have been shown to encode for essential genes in the context of yeast knock-out experiments \cite{Jeong2001} and are also frequently altered in cancer \cite{Hornberg2006,Jonsson2006}, highlighting that topological properties of these networks may carry important biological information. More recently, hierarchical modularity and betweenness centrality are other network properties that have been actively investigated in the context of biological networks \cite{Han2004,Yu2007}. Specifically, the observation that proteins of high betweenness centrality tend to be also essential \cite{Yu2007} and may also be frequently altered in cancer \cite{Ozgur2008,Taylor2009} suggests that proteins of high betweenness centrality are also important for normal cellular physiology. However, it is recognised that well-known disease genes are more frequently studied leading to literature bias and hence to a skew in the topological properties of such genes \cite{Platzer2007}.\\
Existing PINs are also largely incomplete and therefore only represent very crude models of protein interactions \cite{Prasad2009}. Nevertheless they have provided useful structural models and their integration with gene expression data has led to novel insights, also in cancer genomics where they have helped identify candidate gene modules driving disease progression and novel diagnostic/prognostic markers \cite{Chuang2007,Taylor2009}. A distinct advantage of the integrative PIN-mRNA expression approach is that it can overcome some of the inherent problems associated with analysing gene expression data on its own, such as poor signal to noise ratios and the difficulty of interpreting gene expression signatures. Indeed, integrative PIN-mRNA expression studies have helped tease out relevant patterns of expression variation in the contextual framework of signalling pathways and protein complexes \cite{Rapaport2007,Ulitsky2007}. Another potential advantage of using PINs is in relation to expression correlation networks to distinguish between direct and indirect correlations. For example, partial correlations and Graphical Gaussian Models have been proposed for deriving relevance networks where edges represent only potential direct influences between genes \cite{Schaefer2005}. An alternative to partial correlations is to use the structural PIN to remove the (long-distance) indirect correlations between genes.\\
So far though the focus has been on the identification of specific subnetworks that may be causally involved in disease progression \cite{Tuck2006,Chuang2007,Pujana2007,Taylor2009,Nibbe2010}. Much less has been done on the exploration of statistical properties of networks underlying the cancer phenotype with the exception of \cite{Platzer2007} where they performed a comprehensive analysis of cancer networks by first selecting genes differentially expressed between cancer and normal tissue (the ``seeds'') and then constructing the networks as the nearest-neighbor PIN expansion of these seeds. It has become clear however that using PIN information a posteriori does not make full justice of the potential offered by a genuine integrative PIN-mRNA expression analysis. Indeed, it is increasingly recognised that approaches based on single-gene differential expression analysis are too simplistic and have been gradually superseeded by network-based methods \cite{Chuang2007,Taylor2009}. A deeper understanding of the statistical properties of the underlying cancer networks is therefore important, as this may guide the choice of network metrics to use for better identifying signalling pathways or functional modules involved in disease progression.\\
In this work we investigate statistical properties \cite{Barrat2008} of integrated PIN-mRNA expression networks with the aim of identifying network properties that vary significantly between different cancer phenotypes. In particular, we hypothesized that since the frequency of genomic alterations is generally associated with a poor prognosis phenotype \cite{Carter2006,Negrini2010}, that this could be reflected by disruptions in the integrated PIN-mRNA expression networks, which would manifest as increases in the randomness of the local correlation patterns. To this end, we here introduce a local entropy measure that quantifies such randomness and demonstrate that (i) entropy is increased in primary tumours that metastasize (henceforth ``metastatic phenotype''), (ii) that it better characterises the metastatic phenotype than other metrics which do not quantify randomness, and (iii) that this entropy measure can be used to discover genes and associated functional modules involved in cancer progression. To our knowledge this constitutes the first demonstration that metastatic cancer is characterised by a higher degree of randomness in the underlying signal transduction patterns and that this effect can be measured from integration of gene expression data with existing PINs. 

\section*{Methods}

\subsection*{The protein interaction network}
We downloaded the Human Protein Reference Database (HPRD) interaction network \cite{Prasad2009} from Pathway Commons ({\it www.pathwaycommons.org}). The HPRD network has been manually extracted from the literature by expert biologists who read, interpret and analyze the published data. Protein interactions in this network include physical stable interactions such as those defining protein complexes, as well as transient interactions such as post-translational modifications and enzyme reactions found in signal-transduction pathways, including 20 highly curated immune and cancer signalling pathways (``NetPath''). The HPRD network used here consisted of 8396 proteins with 36877 documented interactions.

\subsection*{Breast cancer gene expression data sets}
We downloaded the normalised data of four breast cancer gene expression data sets from GEO ({\it http://www.ncbi.nlm.nih.gov/geo/}) \cite{Wang2005,ChinK2006,Loi2007,Schmidt2008}. These data sets were all profiled with the Affymetrix platform. We refer to these data sets as ``EMC'' (Erasmus Medical Centre) \cite{Wang2005}, ``Frid'' \cite{ChinK2006}, ``LoiUnt'' \cite{Loi2007} and ``Mainz'' \cite{Schmidt2008}. In the case of \cite{Loi2007} we used the data from the untreated patient population. As clinical endpoint we used a binary outcome variable indicating if the tumour metastasized or not. This was possible for the Mainz, Frid and LoiUnt data sets. For the EMC cohort, we used recurrence as a surrogate. The number of samples and their breakup according to clinical outcome in each study were 286 (107 recurred, 179 no recurrence) for ``EMC'', 200 (46 metastasized, 154 no metastasis) for ``Mainz'', 129 (27 metastasized, 102 no metastasis) for ``Frid'' and 125 (28 metastasized, 97 no metastasis) for ``LoiUnt''. To simplify the jargon we also refer to samples that recurred (EMC cohort) as metastatic samples.

\subsection*{Integrated PIN-mRNA expression networks}
Next, we built integrated PIN-mRNA expression networks from the HPRD PIN and a gene expression data matrix. From the PIN we first extracted the proteins for which there were corresponding gene expression profiles. Typically, this resulted in a reduced PIN over approximately 7000 common genes/proteins with a 1-1 correspondence between nodes in the PIN and gene expression profiles. The reduced PIN had several connected components, the maximally connected component containing approximately 95$\%$ of the original nodes. Let $P_{ij}$ denote the adjacency matrix of the PIN and let $N$ denote the number of nodes.\\
Since we are interested in investigating differences in network properties between metastatic and non-metastatic cancer we constructed correlation expression networks by separately computing gene pairwise Pearson correlations across metastatic and non-metastatic cancers. Since typically the number of metastatic samples is less than that of non-metastatic tumours, to ensure that results are independent of system-size, we subselected a number of non-metastatic tumours at random from the original set. Thus, let $C^{(M)}_{ij}$ denote the Pearson correlation of genes $i$ and $j$ across the $n_M$ metastatic samples, and let $C^{(N)}_{ij}$ denote the corresponding correlation across $n_M$ randomly selected non-metastatic samples. We then assign to each pair of genes an edge weight $w_{ij}=C_{ij}$. Finally, since we are interested in the patterns of local correlation we set $w_{ij}=0$ whenever $P_{ij}=0$. Thus, we assign non-zero weights only to edges that are present in the corresponding PIN. The resulting network defines the integrated PIN-mRNA network, and while the structural networks are identical for the metastatic and non-metastic phenotypes, the values of the edge weights will differ.

\subsection*{The stochastic information flux matrix}
There is still freedom in how we can assign weights to the edges in the PIN-mRNA network. In particular, since $-1\leq C_{ij}\leq 1$, we may redefine the edge weights as $C_{ij}\rightarrow w_{ij}\equiv\frac{1}{2}(1+C_{ij})$ for both the metastatic and non-metastatic correlation networks. This monotonic transformation ensures that the edge weights satisfy $0\leq w_{ij}\leq 1$. A weight close to 1 represents therefore a strong positive correlation between the two genes, a weight of approximately 0.5 describes a non-existent correlation, while strong negative correlations will be assigned small weights. For any given node $i$ with neighbours $j\in N(i)$ where $N(i)=\lbrace j\in N : P_{ij}=1\rbrace$, we can then assign a probability distribution as follows:
\begin{equation}
p_{ij}=\frac{w_{ij}}{\sum_{j\in N(i)}{w_{ij}}}
\label{eq:pij}
\end{equation}
so that $\sum_{j\in N(i)}{p_{ij}}=1$. We note that while this defines a stochastic matrix, it is {\it not} doubly stochastic, i.e $\sum_{i\in N(j)}{p_{ij}}\neq 1$, because in general $p_{ij}\neq p_{ji}$.\\
We interpret this probability distribution as follows. Strong positive correlations in the PIN very likely describe interactions that are necessary for the cell to carry out a specific modular function. We call this a ``positive information flux'' interaction to indicate that this interaction describes propagation of an activating signal. Thus, although the positive interaction could be between members of a given complex without explicitly representing signal transduction flow, we nevertheless argue that the positive correlation is necessary for the protein complex to carry out a specific function within the (signal transduction) pathway in which it operates. Because of this, we interpret negative correlations as ``negative information flux'' interactions to indicate that complexes may be in an inactive state or to represent inhibitory interactions. Thus, these interactions are given the smallest ``positive flux'' weights. Weak or near-zero correlations are given a larger positive flux ($p_{ij}\sim 0.5$) than negative interactions ($p_{ij}\sim 0$) because we argue that these interactions have less evidence for inducing an inactivating signal. Thus, the integrated PIN-mRNA networks with the edge weights defined by $p_{ij}$ as in equation (\ref{eq:pij}) describe positive flux interactions (as measured by gene-gene correlations in expression) subject to the structural constraint of the PIN. Applying this procedure to the two $C^{(M)}$ and $C^{(N)}$ correlation matrices, yields two integrated PIN-mRNA networks, one for the metastatic phenotype with stochastic matrix $p^{(M)}_{ij}$, and one for the non-metastatic phenotype with stochastic matrix $p^{(N)}_{ij}$.\\
Since the number of metastatic samples is less than that of non-metastatic tumours (true for all cohorts considered here), and since we constructed the weighted networks using the same number of samples from each phenotype, we have freedom in the choice of non-metastatic samples to use. We exploit this freedom by making in each cohort 10 random selections of non-metastatic samples (in each case selecting the same number of samples as there are of metastatic cases), called ``bootstraps'', which allows the robustness of our results to be evaluated.\\
We point out that gene expression gives only a rough proxy for protein activity and that therefore our integrated PIN-mRNA networks and stochastic information flux matrices only represent very crude models of signal transduction flow. In particular, we expect many egdes to be assigned low fluxes when in fact corresponding protein levels and activities may be highly correlated. Thus, it is clear that our networks do not represent truly integrated networks. However, in the absence of large-scale protein expression and phosphorylation data, specially in relation to cancer phenotypes, relying on correlations in mRNA expression between genes whose protein products have been documented to physically interact (including postranslational modifications) is one viable alternative.

\subsection*{Entropy of information flux distribution}
We propose the following entropy measure \cite{Barrat2008} that quantifies the amount of randomness/disorder in the local flux distribution surrounding any given node $i \in N$. This measure is similar to the local disparity/heterogeneity index of weighted networks considered in other contexts \cite{Derrida1987,Barthelemy2003}. Specifically, if $k_i$ denotes the degree of node $i$, then the local entropy of node $i$ is defined by
\begin{equation}
S_i=-\frac{1}{\log{k_i}}\sum_{j\in N(i)}{p_{ij}\log{p_{ij}}}
\end{equation}
Thus, in the extreme case where all the flux is constrained along one interaction, $p_{ij}=\delta_{ij}$ (where $\delta_{ij}$ is the Kronecker-delta), $S_i=0$, while in the case where the flux is distributed equally across all edges i.e if $p_{ij}=1/k_i$, the entropy attains its maximum value $S_i=1$.\\
We denote by $S$ the distribution of local entropies across the whole network, i.e $S=\{S_1,...,S_N\}$ where $N$ is the number of nodes in the network. Since the stochastic matrix $p_{ij}$ is defined separately for metastatic and non-metastatic phenotypes we obtain separate distributions of local entropies $S^{(M)}$ and $S^{(N)}$.\\
To determine if these two distributions are significantly different we use the non-parametric paired Wilcoxon rank sum test. The paired test is appropriate because we need to control for possible dependencies between the two entropy values associated with each node (see below). We use the P-value of the test as a measure of the degree of difference between the values in the two phenotypes. We also define a differential entropy, $dS$ (also denoted as ``diffS''), on a node-by-node basis, by
\begin{equation}
dS_i=S^{(M)}_i-S^{(N)}_i
\end{equation}
It is important to point out that the sampling properties of the entropy measure as defined here will depend on the degree of the node. To see this, consider a node of degree 10, with half of the neighbors having $p_{ij}=1/5$ while the other half have $p_{ij}=0$. Then, the entropy is $S=\log{5}/\log{10}\approx 0.70$. Consider now another node of degree 100, again with half of the neighbors having $p_{ij}=1/50$ and the rest $p_{ij}=0$. In this case, the entropy is $S=\log{50}/\log{100}\approx 0.85$. Thus, while the two nodes have, modulo the degree, the same stochastic flux distribution, the entropies are widely different with the higher connected node exhibiting much higher entropy. A consequence of this is that the sampling properties and in particular the expected variance of the entropy will decrease with increasing degree. This means therefore that when comparing entropy measures between the two phenotypes, the comparison must be done on a degree-by-degree basis. In particular, the paired Wilcoxon (non-parametric) test used above to evaluate differences in the global entropy distributions performs the comparison on a node-by-node basis and therefore circumvents the degree-dependence problem posed by the entropy measure.

\subsection*{Feature selection based on local differential entropy}
It is clear however that ranking nodes according to the magnitude of entropy change may skew selection to nodes of small degree as high degree nodes are less likely to show big entropy changes. To address this difficulty we adopt an empirical approach to derive a null distribution of expected entropy changes as a function of the degree of the node. Specifically, the approach we used is:
\begin{enumerate}
\item Randomise the gene expression profiles of each sample, reconstruct stochastic flux matrices for the two phenotypes and calculate differential entropy values.
\item For nodes with the same degree, estimate sampling variance. Generally, for nodes of higher degree (i.e $k_i>50$), there were too few nodes to estimate sampling variance, so high-degree nodes were pooled together and a single variance estimate obtained for each pool. In particular, nodes with degrees in the following ranges were pooled in corresponding bins: (50,75), (75,100), (100,150), (150,$\max{k_i}$). This procedure was justified as we observed that the sampling variance was fairly constant for high-degree nodes.
\item Repeat steps 1 \& 2 a total of 100 times to obtain 100 variance estimates for each degree or pooled degree.
\item Obtain smoothed standard error estimates by averaging the standard errors over the 100 randomisations.
\item We observed that the resulting standard deviation, $\sigma$, was a decreasing monotonic function of the degree of the node. In particular, it took the power-law form
\begin{equation}
\sigma(k)=a/k^b
\end{equation}
\item Estimate the parameters $a$ and $b$ using non-linear least squares \cite{Bates1988}.
\item For node $i$ of degree $k_i$ and observed differential entropy value $dS_i$ we computed a P-value by comparing $dS_i$ to the value expected under the null given by a Gaussian of mean 0 and standard deviation $\sigma(k_i)$ with the parameters as estimated in the previous step.
\item Finally, nodes are ranked according to the significance of the P-values, and  subsequent FDR estimation is applied \cite{Storey2003}.
\end{enumerate}

\subsection*{Average local correlation}
We compare our local entropy metric to an alternative metric that quantifies the strength of the local positive flux distribution. Specifically, given a node $i$ we can estimate the strength of the positive information flux surrounding node $i$ by simply averaging the correlations over the nearest neighbors
\begin{equation}
\bar{C}_i=\frac{1}{k_i}\sum_{j\in N(i)}C_{ij}
\end{equation}
Since this is done for each cancer phenotype separately, we obtain for each node the difference in the mean local correlation as
\begin{equation}
d\bar{C}_i=\bar{C}^{(M)}_i-\bar{C}^{(N)}_i
\end{equation}
In addition, we also consider the average of local absolute correlations, that is, 
\begin{equation}
\bar{|C_i|}=\frac{1}{k_i}\sum_{j\in N(i)}|C_{ij}|
\end{equation}
and similarly, the difference in mean local absolute correlation,
\begin{equation}
d\bar{|C_i|}=\bar{|C_i|}^{(M)}-\bar{|C_i|}^{(N)}
\end{equation}
However, since generally we would expect larger local entropies to reflect decreases in the positive information flux or to reflect lower absolute correlations (i.e less order), we redefine these metrics as
\begin{eqnarray*}
\bar{H}_i&=&1-\frac{1}{k_i}\sum_{j\in N(i)}|C_{ij}|\\
\bar{D}_i&=&1-\frac{1}{k_i}\sum_{j\in N(i)}C_{ij} \\
\end{eqnarray*}
and use these metrics in the comparisons to entropy. In comparing all the metrics between the two phenotypes, the P-values obtained from the paired Wilcoxon rank sum tests are directly comparable, since the same number of values are being compared for each choice of metric.

\section*{Results}

\subsection*{Using entropy to characterise the metastatic cancer phenotype}
Based on the observation that more aggressive metastatic breast cancer is associated with a higher genomic instability (i.e a higher frequency of genomic alterations) \cite{Chin2007,ChinK2006,Carter2006,Negrini2010}, we sought to investigate if this could be reflected in integrated PIN-mRNA expression networks. Specifically, we hypothesized that genomic (and epigenomic) alterations would lead to disruptions in the dynamics of information flow on protein interaction networks, and that this disruption would manifest itself as a higher degree of randomness in the patterns of {\it local} gene expression correlations.\\
While gene expression data represents a steady state, we posited that information flux in the network and under a certain condition is driven by positive correlations in gene expression between the interacting proteins. We thus assigned to each edge in the PIN a positive ``flux weight'', the magnitude of which reflects the strength of the Pearson correlation between the corresponding expression profiles (Methods). The weights were normalised further to yield a stochastic matrix $p_{ij}$ ($\sum_{j\in N(i)}p_{ij}=1$ for all $i\in N$, Methods), describing the relative strengths of the local Pearson correlations. Thus, $p_{ij}$ can be interpreted as the probability of signal transmission from protein $i$ to a neighbor $j$ relative to all other interacting neighbors of protein $i$. Using this stochastic matrix we propose to use an entropy-like measure that quantifies the amount of disorder or randomness in the local flux distribution (Methods). Specifically, for each node $i$ we compute an entropy of flux distribution, $S_i$, as
\begin{equation}
S_i=-\frac{1}{\log{k_i}}\sum_{j\in N(i)}{p_{ij}\log{p_{ij}}}
\end{equation}
Thus, nodes with significant changes in $S$ between two phenotypes represent points in the gene network where the degree of randomness in the information flow changes. \\
There are in principle other measures one could use to quantify the disruption in information flow in a network. One possibility is to consider for each node $i$ the distribution of Pearson correlation values $C_{ij}$ with each of its neighbors \cite{Taylor2009}. By comparing the distribution of correlation values $C^{(M)}_i=\{C^{(M)}_{ij}:j\in N(i)\}$ in metastic breast cancer with the corresponding distribution in non-metastatic cancer $C^{(N)}_i$ one can thus identify nodes for which the local flux distribution changes. These nodes may therefore represent genes that are disrupted in the more aggressive phenotype. However, we argued that the degree of randomness in the local flux distribution maybe a better characteristic of a metastatic network than any statistic based on the difference in correlation values ({\bf Figure 1}). Indeed, as the hypothetical example in Figure 1 shows, the metastasis phenotype may be associated with a significant increase in randomness but not necessarily with a significant change in the average nearest-neighbor correlation. On the other hand, there could be nodes with a significant change in the mean nearest-neigbour correlation but without a change in local randomness. We therefore sought to determine if randomness could be a better intrinsic characteristic of the metastatic phenotype than orthogonal measures such as local mean correlation or absolute correlation which do not necessarily quantify such randomness.

\subsection*{Increased entropy in metastatic breast cancer}
To test our hypotheses we first integrated a highly curated network of protein interactions (from the Human Protein Reference Database) \cite{Prasad2009} with a high-quality cancer gene expression data set from 286 breast cancer patients \cite{Wang2005}, of which 107 relapsed (Methods). There were 7279 protein coding genes with expression profiles and with representation in the HPRD PIN. Of these, 167 were isolated nodes, which were therefore removed from the analysis, leaving $n=7112$ nodes. The PIN is sparse containing a total of only 31678 edges, representing 0.01$\%$ of the maximum possible number of edges. The degree distribution exhibited long-tails with the highest degree node having 222 neighbors. About 60$\%$ of nodes had a degree less or equal than 5, while $\sim 75\%$ of nodes had a degree less than 10. To ensure comparibility of correlation values, we sampled 107 samples at random from the 179 non-metastatic samples, to build the corresponding non-metastatic PIN-mRNA network. The metastatic PIN-mRNA network was constructed using all 107 metastatic samples.\\
We first restricted the analysis to the 1903 nodes of degree $\geq 10$, representing about 25$\%$ of all nodes in the network, and compared the entropy measures between the metastatic and non-metastatic networks. First, we found that the local entropies of both metastatic and non-metastatic networks were significantly lower than that of a null network obtained by randomising the expression profile of each tumour sample ({\bf Figure 2A}). This is in line with the expectation that local correlation patterns are far from random. We also observed that our proposed entropy measure was discriminatory of the two cancer phenotypes, with a small yet significant increase in entropy in the metastatic network (Wilcoxon paired test $P=2\times 10^{-16}$). To test the robustness of these results we compared the metastatic network to other non-metastatic PIN-mRNA networks obtained by randomly selecting a different set (``bootstraps'') of 107 non-metastatic tumours. Viewing these as perturbations of the original non-metastatic network, we performed a total of 10 bootstraps and in all cases the entropy of the metastatic network was higher ({\bf Figure 2B}, Wilcoxon paired test $P<0.005$ in all cases). We also observed that differential entropy values were highly correlated between the different bootstraps (mean Pearson correlation 0.77), thus allowing values to be averaged.\\
Differences in local entropy between the metastatic and non-metastatic network were not substantial in absolute terms, however, the paired test results clearly showed statistical significance and a trend towards increased entropy. To obtain a handle on the statistical significance of individual entropy changes, we first noted that the variance in differential entropy values was dependent on the degree of the node ({\bf Additional file 1A}), which was not unexpected (Methods). Specifically, high degree nodes are generally less likely to exhibit large entropy changes. To overcome this problem we derived degree-dependent variance estimates of the differential entropy, which allowed specific P-values to be derived (Methods). We observed that the variance decreased with the degree of the node in the form of a power-law allowing P-values to be estimated for nodes of any degree ({\bf Figure 2C}). Confirming the paired test analysis, individual P-values exhibited a clear skew towards 0 suggesting many significant changes ({\bf Figure 2D}). Among the top 200 ranked nodes (stringent FDR$<0.001$) ({\bf Additional file 2}), 133 showed entropy increases (Binomial test $P<10^{-6}$). Using the P-values to rank the genes also confirmed that the selected nodes were now not skewed towards those of low-degree ({\bf Additional file 1B}). \\
Because these results were obtained on the selected nodes of high degree ($k_i\geq 10$), we asked if this result would remain significant had we included all nodes. We thus repeated the analysis using all 5592 nodes with at least two neighbors (entropy undefined for $k_i<2$). This showed that entropy was also significantly higher in metastatic breast cancer (Wilcoxon paired test $P<10^{-100}$), despite the fact that non-hubs made up 75$\%$ of the network. Thus, while nodes of lower degree generally also showed increases in entropy we focused our analysis on the higher-degree nodes ($k_i\geq 10$) as these are more prone to disruption in cancer \cite{Hornberg2006,Jonsson2006}.

\subsection*{Validation in other breast cancer cohorts}
We next asked if the increase in local entropy found in the EMC data set is a robust finding independent of the expression data set used. We thus collected three additional breast cancer expression data sets that had been profiled with Affymetrix arrays \cite{ChinK2006,Loi2007,Schmidt2008}. For all three data sets we constructed the corresponding PIN-mRNA networks in both metastatic and non-metastatic disease using the same procedure as for the EMC data. The PIN-mRNA networks were of similar size to the one obtained from the EMC data set. Remarkably, as with the EMC PIN-mRNA networks, we observed that the local entropies (of nodes with degree $k_i\geq 10$) were significantly increased in the metastatic networks ({\bf Figure 3A}). In all cases, we verified that the increase in entropy was robust to the choice of non-metastatic network. We emphasize here again that metastatic and non-metastatic PIN-mRNA networks were derived using the same number of samples, thus excluding the possibility that the increase in entropy is a sample-size effect. However, the increase in entropy was less marked when the analysis was extended to include all nodes of degree $\geq 2$ (Mainz $P=4\times 10^{-7}$, Frid $P>0.5$, LoiUnt $P=2\times 10^{-5}$).\\
It is natural to ask if the increase in entropy seen across all cohorts is due to the same genes or instead if in each cohort the subset of genes exhibiting increases is different. To address this we selected the 133 genes which showed the most significant increases in entropy in the EMC data set (out of the top 200 genes with most significant changes ($FDR<0.001$), 133 showed increases) and verified that these generally also showed more significant increases in entropy in the validation sets ({\bf Figure 3B}). This supports the view that there is gene-wise consistency in the metastasis-associated entropy increases seen across all four cohorts. 

\subsection*{Randomness as an intrinsic property of the metastatic phenotype}
Given the robust finding of increased entropy in metastatic breast cancer, we next asked if other local measures could discriminate between the two cancer phenotypes. Specifically, we compared entropy against the average local correlation, that is, the mean correlation of a node with its nearest neighbors, as well as the mean local absolute correlation (Methods).\\
We expected that the mean correlation (or absolute correlation) in expression would decrease in metastatic breast cancer as a result of the increased frequency in genomic/epigenomic alterations. We confirmed this across all four data sets using the same one-tailed paired Wilcoxon test ({\bf Table 1}). Comparison of the P-values obtained using entropy and average correlation measures however also showed that entropy was substantially more different between metastatic and non-metastatic breast cancer ({\bf Table 1}).\\
As an alternative test we also asked how many nodes exhibited an increase or decrease in entropy, average correlation or absolute correlation and tested the strength of any deviation from the null using the binomial test. By performing this binomial test we are also comparing entropy to a paired t-test of correlation values as the directionality of change is determined by the difference in mean correlations. We observed that there were a substantially larger number of nodes exhibiting increased entropy in metastatic breast cancer than nodes exhibiting lower average correlation ({\bf Figure 4A}). In line with this, binomial test P-values of skewness derived from differential entropy values were also substantially more significant than those derived from differences in mean local correlation ({\bf Figure 4B}). Interestingly, comparing differential entropy to changes in the local mean absolute correlation, we observed that the absolute correlation was a good surrogate for entropy. However, entropy did generally show more significant changes than the absolute correlation ({\bf Table 1, Figure 4}). Thus, these results demonstrate that increased entropy in the local patterns of gene expression correlations may be a more intrinsic property of the metastatic phenotype and thus constitute a more useful measure for identifying gene subnetworks implicated in breast cancer metastasis than subnetworks derived from direct correlation measures.

\subsection*{Biological significance}
We would expect that nodes exhibiting the most significant increases in entropy may point to genes and signalling pathways that are more frequently altered in the metastatic or poor prognosis breast cancer phenotype, irrespective of the modality of the underlying alteration. Of the 133 genes showing the most significant increases in entropy {\bf Additional file 2}) several have expression levels that have already been associated with a poor prognosis and distant metastasis in breast cancer ({\it e.g CDC2, CCNB1, MYBL2, MAD2L1}) and which have functions related to the cell-cycle and DNA replication \cite{Teschendorff2006,Sotiriou2006}. Graphical depiction of the nearest neighbor subnetwork surrounding e.g {\it MYBL2} confirmed the increase in entropy in the metastatic phenotype ({\bf Figure 5}). For instance, one can observe how a strong inhibitory interaction with {\it CCND1} and strong correlations with {\it CCNE1} and {\it RBL1} become disrupted in the metastatic phenotype. It is clear that nearest neighbors of {\it MYBL2} (e.g {\it CCNA2}) also undergo increases in entropy, thus implicating larger gene modules, in this case a cell-cycle module, that are altered in metastatic disease. Interestingly, GSEA (Gene Set Enrichment Analysis using a Fisher-test) of genes with increased entropy in the metastatic phenotype revealed enrichment of several biological pathways, notably apoptosis, natural killer cell mediated cytotoxicity and interleukin-2 (IL2), androgen receptor and insulin growth factor (IGF) signalling ({\bf Table 2}).  Corresponding subnetworks for {\it IL2RB}, {\it IGFBP7} and {\it BCL2} clearly confirmed the increase in entropy associated with these genes and pathways ({\bf Figure 5}).\\
Since GSEA was performed on the top 200 genes using the 1903 gene list as the background reference, the enrichment of any biological pathways and in particular those of apoptosis and IL2/IGF is very significant and not caused by the selection of nodes of high degree. As a control and to confirm this further, we observed no enrichment of any biological pathway among the genes showing decreases in entropy ({\bf Table 2}). Although this could be explained in part by the smaller number of genes showing decreases (67 vs 133), we only obtained significance of one biological term (regulation of translation initiation) when we expanded the 67 gene set to the 133 with the most significant entropy decreases.\\
It is likely that the increases in entropy associated with specific genes are caused by a higher frequency of underlying mutations, losses or epigenetic silencing. Using copy-number and sequencing data of a large cohort of 171 breast tumours (38 metastasized, 130 no-metastasis) \cite{Chin2007} we observed an increased frequency of losses in those tumours that metastasized for many of the top ranked genes including {\it BCL2} (18$\%$ vs 13$\%$), {\it IGFBP7} (11$\%$ vs 6$\%$), {\it IGF1} (8$\%$ vs 2$\%$) ({\bf Additional file 3}). Similarly, {\it TP53} was mutated in 31$\%$ of tumours that metastasized against 17$\%$ in those tumours that did not. Of the top 133 genes exhibiting increases in entropy, 90 could be mapped to the array CGH study and of these, 68 (75$\%$) showed more frequent losses in primary tumours that metastasized ({\bf Additional file 4}). In contrast, 65$\%$ of genes exhibiting no significant changes in entropy were more frequently lost in metastatic cases ({\bf Additional file 4}). Thus, genes with significant increases in entropy were approximately 1.6 times more likely to undergo more frequent losses in disseminating tumours (Fisher test $P=0.08$).

\section*{Discussion}
In this work we have explored a specific statistical property of integrated protein interaction and
cancer mRNA expression networks. Our hypothesis was that the metastatic cancer phenotype is characterised by a higher level of randomness in the sense of a more disordered local gene expression correlation pattern. In other words, since one may view the protein interactions as imposing to some degree constraints on the allowed local gene expression patterns \cite{Yu2007,Taylor2009}, we posited that these constraints would be violated to a higher extent in breast tumours which exhibit a more invasive phenotype. To test this we introduced a stochastic matrix on the PIN, modeling positive information fluxes around any given node, and from it we defined a local entropy measure which quantifies the degree of randomness of this flux distribution. Using this entropy measure we could show that the metastatic phenotype is indeed characterised by a higher level of entropy independently of the breast cancer cohort considered. While entropy changes were not large in absolute terms due to the focus on nodes of relatively high degree (degree $\geq 10$), we were able to show using an empirically derived degree dependent variance estimator, that the changes in entropy were significant with a clear skew towards higher entropy in the metastatic phenotype. Moreover, we were able to show that genes exhibiting significant increases in local entropy in one data set, did so also in the independent cohorts, supporting the view that the disruptions in local expression patterns that we have found are biologically genuine and of relevance to the metastatic phenotype. It is interesting though that the increase in entropy was less marked when analysis was extended to all nodes of degree larger than 1, indicating that low-degree nodes showed less consistent directional changes, in line with the expectation that low-degree nodes may be functionally less important. \\
We also observed significant enrichment of relevant biological pathways, specially those with tumour suppressor functions, among genes showing increases in entropy in the metastatic phenotype, but not so for genes showing decreases in entropy. This supports the view that it is increases in randomness that may be of most functional consequence. A higher randomness in the coexpression patterns of genes important to the apoptotic cascade may indicate that this pathway is less functional in primary tumours that eventually metastasize. Similarly, our findings suggest that increased randomness in the patterns of expression of the IL2 pathway, a well-known tumour suppressor pathway mediating tumour inhibition through formation of natural killer cells \cite{Pardoll2009}, is a critical determinant of breast cancer metastasis. This confirms other reports that immune-response pathways are important in prognosis prediction of breast cancer \cite{Teschendorff2007gb,Teschendorff2008}.  While the IGF-pathway has been implicated as a mediator of breast cancer progression and is also involved in the regulation of apoptosis \cite{Neuberg1997,Samani2007,Yee2008}, our results suggest a novel important role for {\it IGFBP7} and IGF-signalling.\\
Generally, there were other instances where the identified genes formed interlinked subnetworks implicating whole gene modules that show increased entropy. It will therefore be interesting to extend our analysis to metrics that quantify the degree of randomness across whole subnetworks as opposed to single nodes. Among the larger modules, we identified one associated with the cell-cycle ({\it CDC2, CDC20, MYBL2, MAD2L1}) consistent with many other studies showing the importance of cell-cycle genes in breast cancer prognosis \cite{Teschendorff2006,Sotiriou2006}. Of note, we also found local entropy increases in subnetworks surrounding genes involved in invasion ({\it SPARC, MMP1, MMP2, MMP3}). {\it SPARC} expression itself has been associated with clinical outcome in breast cancer in a large number of studies, for example see \cite{Naderi2007,Farmer2009}. In this regard, it is worth pointing out that we have rediscovered these important cell-cycle and extracellular matrix modules without directly comparing expression
levels between metastatic and non-metastatic breast cancer, but rather by using entropy to compare the randomness of expression patterns within each phenotype.  Thus, our proposed entropy metric is able to identify biological mechanisms related to the cancer hallmarks and a poor prognosis phenotype.\\
Perhaps most remarkably is the fact that entropy outperformed other local measures of information flux, in the sense that entropy better characterised the metastatic phenotype. One competing measure we considered was the average nearest-neighbor correlation in gene expression, which clearly was less effective in distinguishing the two cancer phenotypes. Similarly, entropy characterised the metastatic phenotype more closely than the effect size statistic provided by a t-test. The t-test is similar to the method used in \cite{Taylor2009} to identify features disrupted in poor prognosis breast cancer. In that study, the authors identified {\it BRCA1} as one of the top genes exhibiting changes in the mean local correlation. In terms of differential entropy however, {\it BRCA1} was not highly ranked because differential entropy only measures the change in the degree of randomness of the local correlation distribution, while as shown in \cite{Taylor2009} {\it BRCA1} exhibits a fairly uniform switch from positive to negative correlations with its interacting partners, a switch which is therefore not necessarily associated with an increase in disorder or randomness. This example highlights the fact that although differential entropy can miss important genes, that it constitutes a metric that is complementary to the one used in \cite{Taylor2009}. Therefore, it will be interesting to compare the alternative metrics in relation to the ranking of nodes. However, the result obtained here that increased entropy is a better distinguishing feature of the metastatic phenotype supports the view that changes in the randomness of the local correlation patterns may be more relevant than changes in mean correlation levels. To confirm this further, we took the mean of the absolute correlations as another metric and compared it to entropy. In line with the expectation that increases in entropy are generally caused by drops in absolute correlation values, we observed that changes in the mean absolute correlation, while not outperforming entropy, was a good surrogate for it. It will therefore also be interesting to compare our entropy metric to other network metrics that quantify randomness or to metrics that combine randomness and changes in mean correlation levels.\\
The observed increase in entropy in the metastatic breast cancer networks may have a clear biological interpretation, as genomic and epigenomic alterations are more frequent in primary tumours that metastasize \cite{Carter2006,ChinK2006,Chin2007}. This increased frequency of alterations is true for losses, mutations and amplifications (high-level gains) but does not necessarily hold for low-level gains \cite{ChinK2006,Chin2007}. It was therefore encouraging to observe that genes undergoing the most significant increases in entropy were also more likely to undergo more frequent losses in the tumours that metastasized. Although this association was only marginal, this could be because the genomic alteration data came from a different cohort. Another potential difficulty in linking entropy changes to gene alterations is that a substantial number of alterations are not copy-number changes but instead may represent point mutations, rearrangements, or epigenetic aberrations, and large scale profiling of these types of alterations is still ongoing. Thus, the question if differential entropy selects for genes that are more frequently altered and how the directionality of change might relate to the type of alteration is still an open question that we hope to address in future studies. An easier context in which to study this would be to compare integrated PIN-mRNA networks for cancer and corresponding normal tissue, as normal tissue exhibits significantly much lower levels of alterations. However, this also requires relatively large normal and cancer tissue data sets profiled on the same platform and ideally as part of the same study. While in breast cancer such data sets are only now becoming available, it will be interesting in the meantime to pursue this question in other cancers.

\section*{Conclusions}
Metastatic breast cancer is characterised by an increase in the randomness of the local expression correlation patterns. The entropy metric proposed here may therefore be a useful tool for identifying genes and signalling pathways implicated in the metastatic process of other cancers. We foresee that further network-theoretical studies of integrated PIN-mRNA expression networks in the context of cancer genomics will be fruitful.

\subsection*{Author contributions}
AET designed, analysed and wrote the paper. SS contributed with ideas and to the writing of the manuscript. All authors read and approved the final manuscript.

\subsection*{Acknowledgements}
AET was supported by a Heller Research Fellowship. SS was supported by a Newton International Fellowship.


\bibliographystyle{bmc_article}  
\bibliography{entropyXXXfv}     



\newpage
\section*{Additional Files}

\subsection*{Additional file 1 - Differential entropy and node degree}
{\bf A)} Differential entropy change is plotted against node degree. One observes a skew in the sense that high degree nodes exhibit smaller changes in entropy. This is a theoretical consequence of the entropy defintion. Thus using $|dS|$ to rank genes is biased to nodes of low degree. {\bf B)} The negatiive of the log(P-values) of differential entropies against node degree. Evaluating the statistical significance of the entropy changes we now observe that for every degree there are significant changes in entropy, thus removing the skew. Green line is line defined by $P=0.05$.

\subsection*{Additional file 2 - Genes ranked according to differential entropy}
All nodes of degree $\geq 10$ are ranked according to significance of differential entropy changes between metastatic and non-metastatic breast cancer (EMC data set). Columns label common gene symbol, the difference in entropy (metastatic minus non-metastatic, averaged over the 10 bootstraps), the degree of the node in the PIN, the estimated P-value and false discovery rate (FDR).

\subsection*{Additional file 3 - Relation of differential entropy to patterns of differential loss and gain in breast tumours}
For the top 200 genes showing the most significant entropy changes, we provide the frequencies of genomic copy-number gain and loss across 171 breast tumours \cite{Chin2007}, stratified according to whether primary tumours metastasized (DM) or not (NoDM). The differential gains and losses are also provided in table. NA indicates that that gene had no oligo representation on the array.

\subsection*{Additional file 4 - Increased entropy and genomic loss}
Of the top 200 nodes with most signficant entropy changes, 133 showed significant increases in the metastatic phenotype, and 90 of these genes could be mapped to an oligo array comparative genomic hybridisation study over 171 breast tumours \cite{Chin2007}. Left panel plots the difference in the frequency of loss of the gene between the tumours that metastasized and those that did not (y-axis) against the negative logarithm of the P-value for the differential entropy (x-axis). Right panel shows the same plot for the 133 genes showing least significant changes in entropy. Genes with significant increases in entropy were 1.6 times more likely to be more frequently lost in poor prognosis tumours (Fisher test $P=0.08$).

\section*{Tables}
\begin{table}[ht]
\begin{center}
\begin{tabular}{|c|cccc|}
  \hline
Metric & EMC ($n=1903$) & Mainz ($n=1900$) & Frid ($n=1270$) & LoiUnt ($n=1899$) \\
\hline
\hline
$d\bar{C}$ & 0.001 & 0.01 & 0.54 & 2e-13 \\
$d\bar{|C|}$ & 1e-16 & 5e-10 & 1e-12 & 3e-70 \\
$dS$       & 2e-16 & 6e-13 & $<$1e-100 & $<$1e-100 \\
\hline
\end{tabular}
\caption{One tailed paired Wilcoxon test P-values comparing distribution of local measures of disruption in information flow in metastatic and non-metastatic PIN-mRNA networks across four different breast cancer cohorts. $d\bar{C}$ denotes the difference in mean local correlation, $d\bar{|C|}$ denotes the difference in mean local absolute correlation and $dS$ denotes the differential local entropy. In the case of the non-metastatic networks, values were averaged over 10 distinct bootstraps before computation of the P-values. The number of pairs (nodes) in the test, $n$, corresponding to the number of nodes in the network with degree $\geq 10$ are given.}
\label{tab:Table1}
\end{center}
\end{table}

\begin{table}[ht]
\begin{center}
\begin{tabular}{|c|ccc|}
  \hline
Pathway & P (dS$>0$) & P(dS$<0$) & Example genes \\
\hline
\hline
Apoptosis & 6e-7  & n.s  & {\it FAS, TP53, TP53BP2, BCL2, BCL2L1, CASP3} \\
IL2 & 6e-4 & n.s & {\it IL2RB, HRAS, FOS, SOS1, LCH, SHC1} \\
AR   & 8e-4 & n.s & {\it SNX1, RBX1, BSG, NOTCH1, RAB27A, TAF9} \\
IGF1 & 9e-4 & n.s & {\it IGF1, IRS1, IGFBP7} \\
\hline
\end{tabular}
\caption{Gene Set Enrichment Analysis of the top 200 genes of which 133 showed increases in entropy. While we observed enrichment of biological pathways among genes showing increases in entropy, there was none among the 67 genes showing decreases. P-values of enrichment (one tailed Fisher's exact test) against genes showing entropy increases (dS$>0$) and decreases are given (dS$<0$) and were calculated using all nodes of degree $\geq 10$ as reference (1903 genes) to avoid intrinsic literature bias. n.s=not significant} 
\label{tab:Table2}
\end{center}
\end{table}

\section*{Figures}

\begin{figure}[h]
\begin{center}
\scalebox{0.80}{\includegraphics{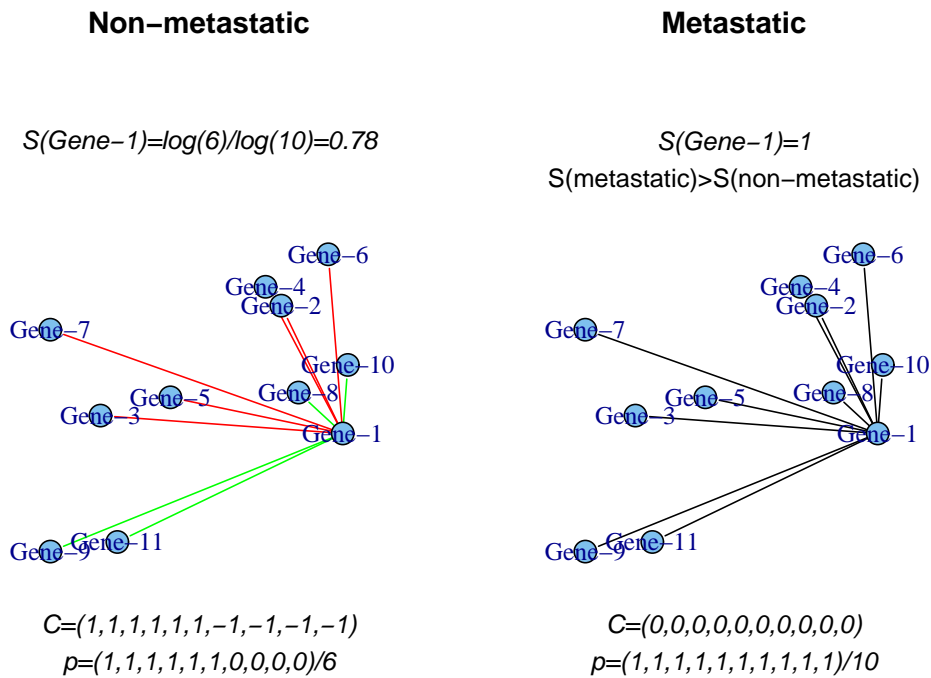}}
\caption{\label{fig:Fig1}Local entropy changes in integrated PIN-mRNA networks:
In the above hypothetical networks, an edge represents a documented interaction between the corresponding proteins in the PIN. The color of the edge codes for the pairwise correlation in mRNA expression in two different conditions, here non-metastatic and metastatic cancer. In this hypothetical example, gene-1 is positively correlated with maximal value ($C=1$) to the first 6 genes, but negatively correlated ($C=-1$) to the remaining four in the non-metastatic PIN-mRNA network. These values translate to a stochastic positive flux vector $p=\frac{1}{6}(1,1,1,1,1,1,0,0,0,0)$ and to an entropy value of 0.78 (Methods). In this hypothetical example, gene-1 is lost/mutated in metastatic cancer, leading to a loss of mRNA expression correlation and anti-correlation with the nearest neighbors. This introduces more disorder/randomness in the local flux distribution, illustrated here by zero correlation values. The resulting local entropy takes on a maximum value of 1, and so for this node there is a large increase in the local entropy which is statistically significant. In contrast, a Wilcoxon rank sum test between the correlation values $C=(1,1,1,1,1,1,-1,-1,-1,-1)$ and $C=(0,0,0,0,0,0,0,0,0,0)$ yields a non-significant P-value of 0.2.}
\end{center}
\end{figure}

\begin{figure}[h]
\begin{center}
\scalebox{0.80}{\includegraphics{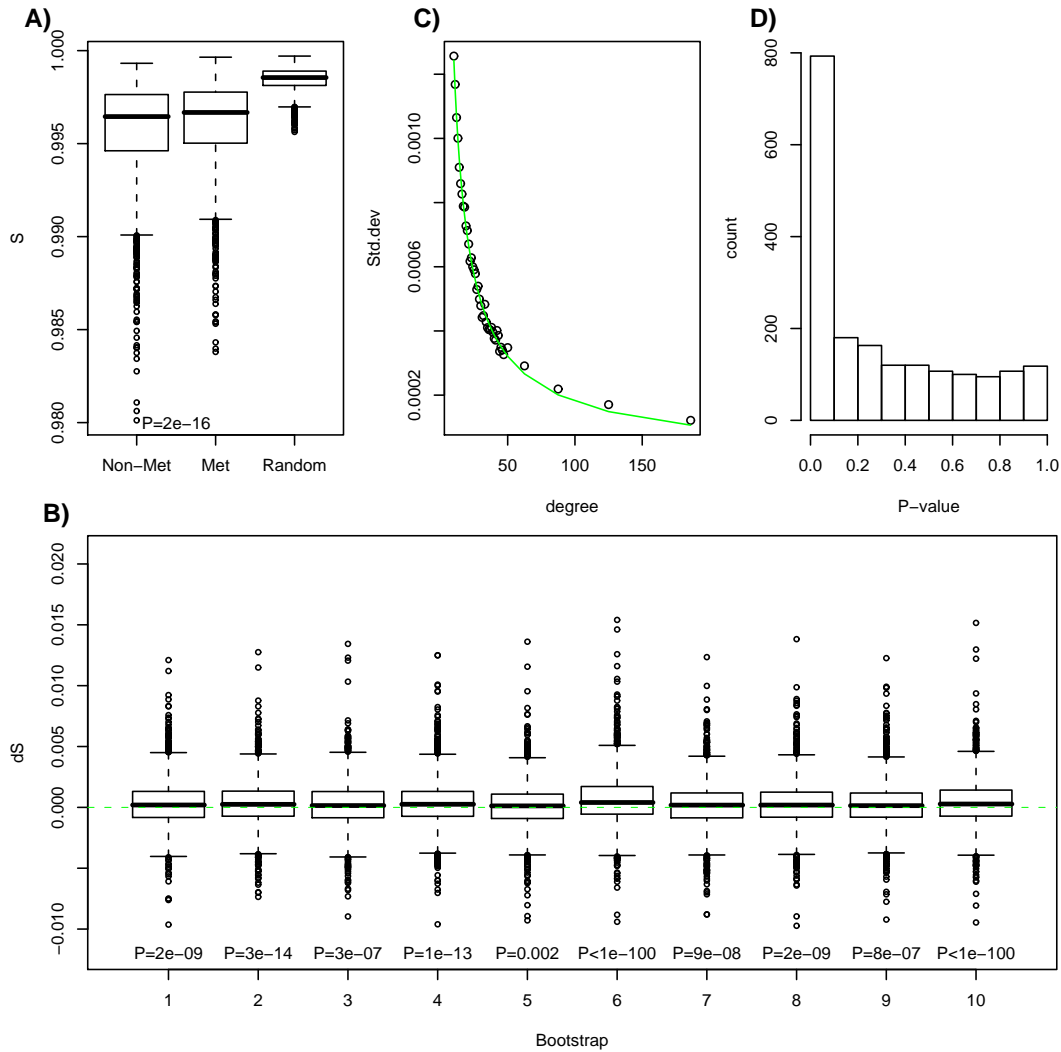}}
\caption{\label{fig:Fig2} Local entropy is increased in metastatic breast cancer: {\bf A)} Comparison of local entropies for the 1903 proteins with degrees $\geq 10$ between the non-metastatic, metastatic and random weighted networks. The entropies in the metastatic network exhibit significantly higher values than those in the non-metastatic network: P-value given is from a one-tailed paired Wilcoxon rank sum test. Both non-metastatic and metastatic networks show significanly lower entropies than those of a purely random network obtained by randomisation of expression profiles. {\bf B)} Differential entropy values (metastatic minus non-metastatic) are significantly greater than zero for 10 different choices of non-metastatic networks obtained by bootstrapping samples. One-tailed paired Wilcoxon rank sum test P-values are given. {\bf C)} The expected variation in differential entropy under the null distribution against node degree. The green-line is a non-linear least squares fit of a power-law function of the form $a/k^b$ where $k$ is the node degree. Estimated parameter values are $\hat{a}=0.0086,\hat{b}=0.08411$. {\bf D)} Histogram of P-values of genes (nodes). P-values were estimated by comparing observed differential entropy values to those expected under the null using the variance estimates from C).}
\end{center}
\end{figure}

\begin{figure}[h]
\begin{center}
\scalebox{0.80}{\includegraphics{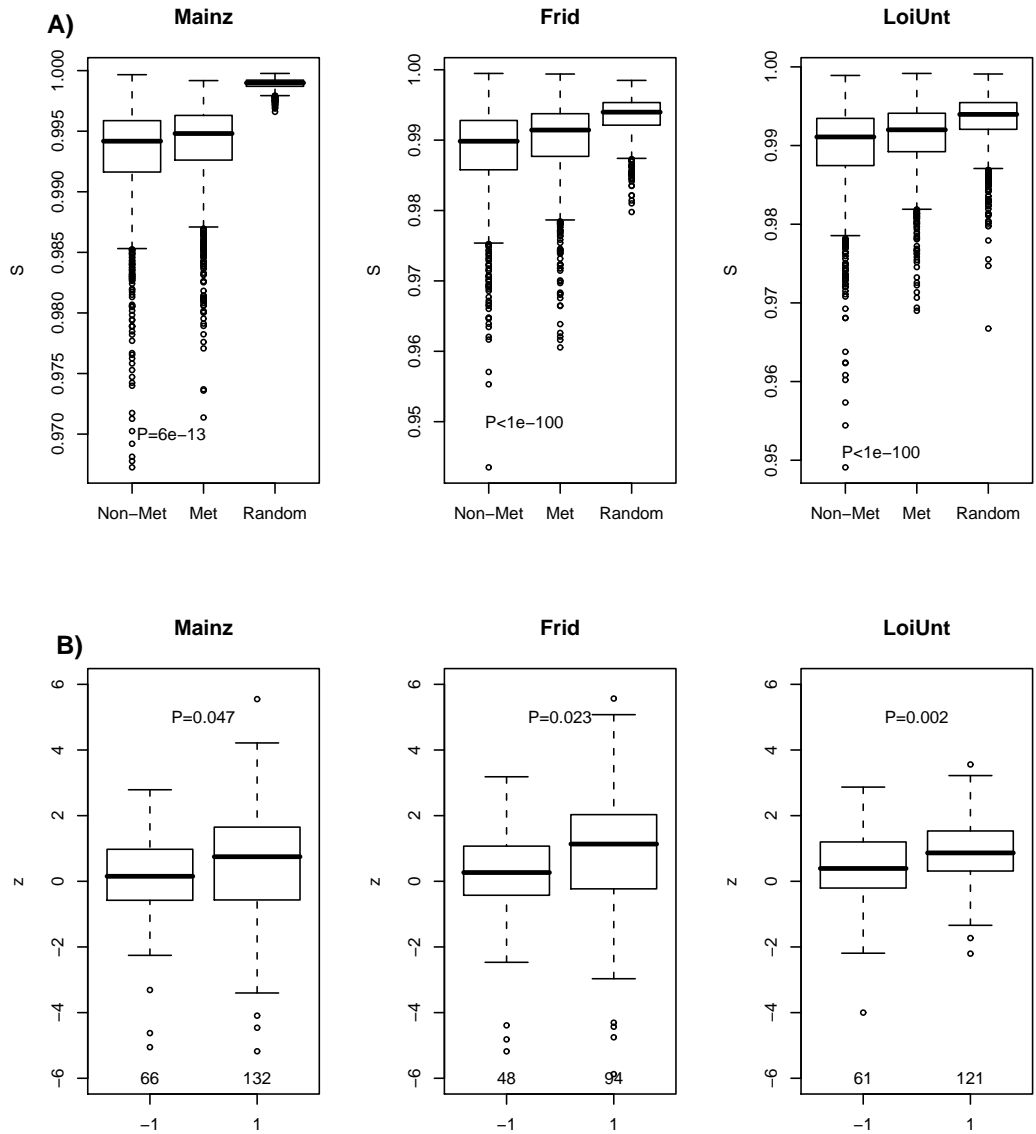}}
\caption{\label{fig:Fig3} Increased entropy in metastatic breast cancer is independent of breast cancer cohort: {\bf A)} Comparison of local entropies for the proteins with degrees $\geq 10$ between the non-metastatic, metastatic and random weighted networks in the three validation cohorts: Mainz (1900 genes), Frid (1270 genes) and LoiUnt (1899 genes). The entropies in the metastatic network exhibit significantly higher values than those in the non-metastatic network: P-value given is from a one-tailed paired Wilcoxon rank sum test. Both non-metastatic and metastatic networks show significanly lower entropies than those of a purely random network obtained by randomisation of expression profiles. {\bf B)} The top 200 genes with most significant P-values (FDR$<0.001$) as identified in the EMC (discovery) data set are grouped according to increases in entropy (1) or decreases (-1). The numbers of genes within each group and represented in the validation set are given just above x-axis. The y-axis labels the statistics ($dS_i/\sigma_i$) of these genes in the corresponding validation set, and the boxplot allows a direct comparison to be made. The P-value is from a one-tailed Wilcox rank sum test as the alternative hypothesis is that the statistics should be larger for those genes identified to undergo increases in entropy in the discovery set.}
\end{center}
\end{figure}

\begin{figure}[h]
\begin{center}
\scalebox{0.80}{\includegraphics{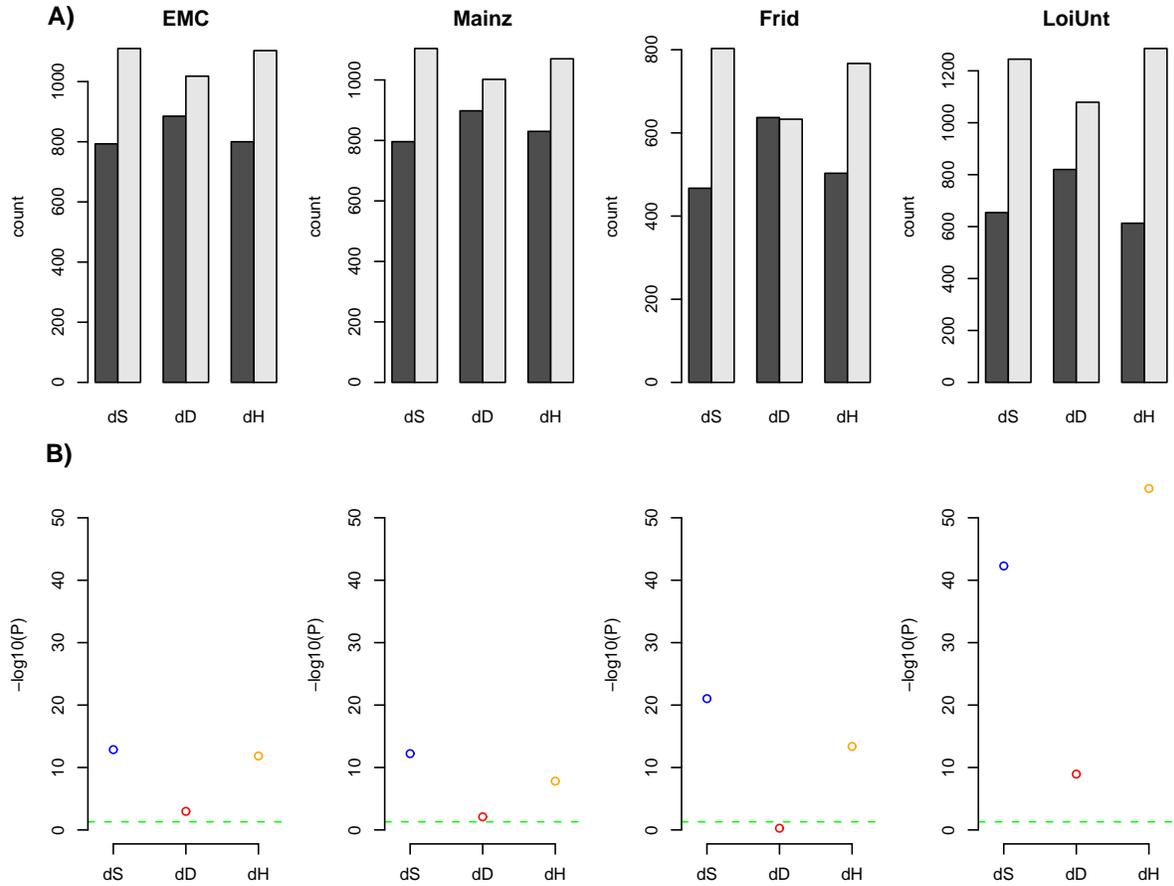}}
\caption{\label{fig:Fig4} Local entropy better characterises the metastatic network: {\bf A)} For each of the four breast cancer cohorts, we count the number of genes showing increases (grey) and decreases (black) in local entropy ($S$), negative local mean correlation ($D$), and heterogeneity (negative local mean absolute correlation) $H$ in the metastatic PIN-mRNA networks. In all cases, nodes of degree $\geq 10$ were selected. {\bf B)} Corresponding -log10(p-values) from a one-tailed binomial test. The line -log10(0.05) is shown in green.}
\end{center}
\end{figure}

\begin{figure}[h]
\begin{center}
\scalebox{0.7}{\includegraphics{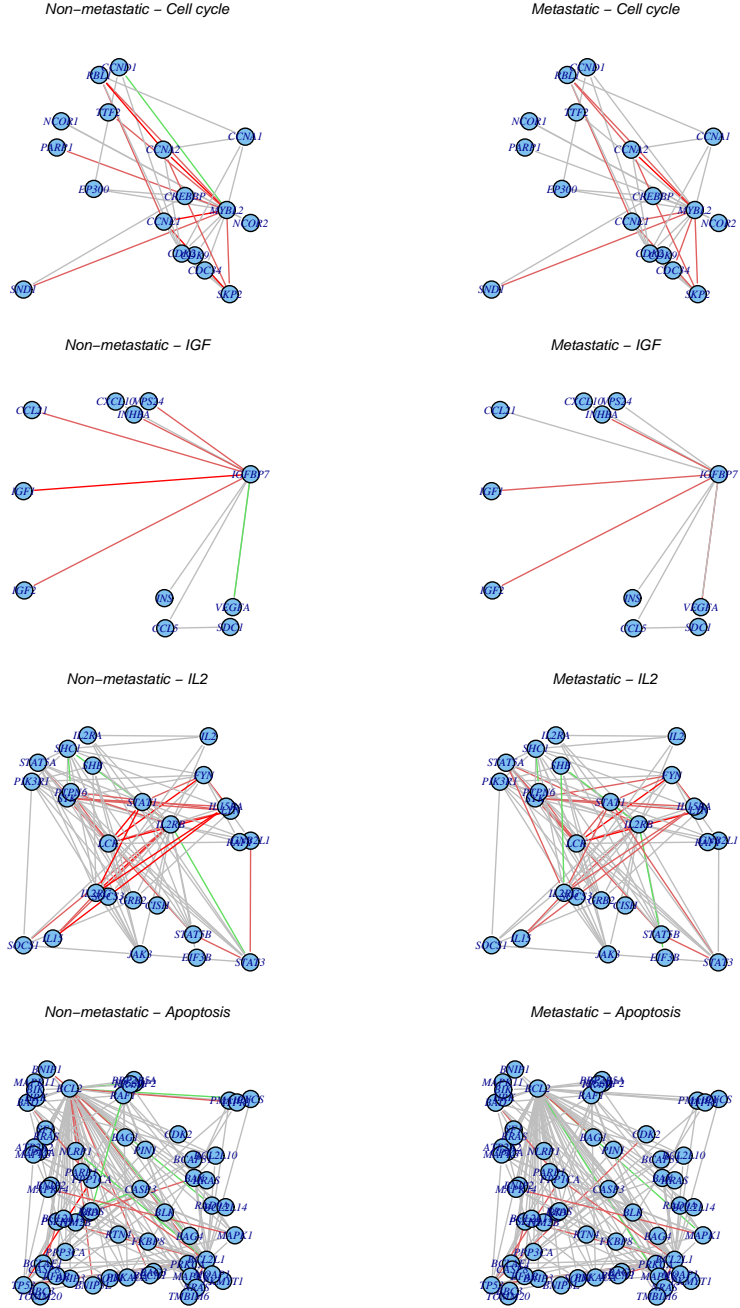}}
\caption{\label{fig:Fig5} Biological subnetworks exhibiting significant increases in entropy: We contrast the integrated PIN-mRNA metastatic and non-metastatic networks for four nodes/genes exhibiting significant increases in entropy and related to tumour suppressor pathways and cancer hallmarks: {\it MYBL2} and cell-cycle, {\it IGFBP7} and IGF-signalling, {\it IL2RB} and IL2 immune-mediated tumour suppression, {\it BCL2} and apoptosis. In each case, we only depict the nearest neighbors (interacting protein partners) of the selected nodes. The edge color shows the strength of the Pearson correlation $C$ in expression between the two genes across the given phenotype: bright red ($0.5<C<1$), dark red ($0.25<C<0.5$), grey ($-0.25<C<0.25$), dark green ($-0.5<C<-0.25$), bright green ($-1<C<-0.5$).}
\end{center}
\end{figure}

\end{document}